\documentclass[amssymb, amsmath, aps, showpacs, superscriptaddress, nofootinbib, twocolumn]{revtex4} %% prb
\usepackage{bm,url,graphicx,natbib}%
\long\def\symbolfootnote[#1]#2{\begingroup\def\thefootnote{\fnsymbol{footnote}}\footnote[#1]{#2}\endgroup}
\newcommand{\degr}{\ensuremath{^{\circ}}\ }

\newcommand{\Vgm}{V_{\rm{g}}}
\newcommand{\Vg}{$V_{\rm{g}}$\ }

\newcommand{\cmVsn}{cm$^2$V$^{-1}$s$^{-1}$}
\newcommand{\DT}{$\Delta T$\ }
\newcommand{\DTm}{\Delta T}
\newcommand{\DTonT}{$\Delta T/T$\ }
\newcommand{\DTonTn}{$\Delta T/T$}
\newcommand{\DTonTm}{\Delta T/T}
\newcommand{\T}{$T$\ }

\begin{document}
\title{Charge trapping in polymer transistors probed by terahertz spectroscopy and scanning probe potentiometry}
\author{J.~Lloyd-Hughes}\email{james.lloyd-hughes@physics.ox.ac.uk} \affiliation{University of Oxford, Department of Physics, Clarendon Laboratory, Parks
Road, Oxford, OX1 3PU, United Kingdom}

\author{T.~Richards} \affiliation{University of Cambridge, Department of Physics, Cavendish Laboratory,
Madingley Road, Cambridge, CB3 0HE, United Kingdom}

\author{H.~Sirringhaus} \affiliation{University of Cambridge, Department of Physics, Cavendish Laboratory,
Madingley Road, Cambridge, CB3 0HE, United Kingdom}

\author{E.~Castro-Camus} \affiliation{University of Oxford, Department of Physics, Clarendon Laboratory, Parks
Road, Oxford, OX1 3PU, United Kingdom}

\author{L.M.~Herz} \affiliation{University of Oxford, Department of Physics, Clarendon Laboratory, Parks
Road, Oxford, OX1 3PU, United Kingdom}

\author{M.B.~Johnston}\email{m.johnston@physics.ox.ac.uk} \affiliation{University of Oxford, Department of Physics, Clarendon Laboratory, Parks
Road, Oxford, OX1 3PU, United Kingdom}

\date{\today}

\begin{abstract}
{Terahertz time-domain spectroscopy and scanning probe potentiometry were used to
investigate charge trapping in polymer field-effect transistors fabricated on a silicon
gate. The hole density in the transistor channel was determined from the reduction in the
transmitted terahertz radiation under an applied gate voltage. Prolonged device operation
creates an exponential decay in the differential terahertz transmission, compatible with
an increase in the density of trapped holes in the polymer channel. Taken in combination
with scanning probe potentiometry measurements, these results indicate that device
degradation is largely a consequence of hole trapping, rather than of changes to the
mobility of free holes in the polymer.}
\end{abstract}

%\pacs{78.20.-e, 78.47.+p}

%78.20.-e    Optical properties of bulk materials and thin films
%78.47.+p    Time-resolved optical spectroscopies and other ultrafast optical measurements in condensed matter

\maketitle

%%%%%%%%%%%%%%%%%%%%%%%%%%%%%%%%%%%%%%%%%%%%%%%%%%%%%%%%%%%%%%%%%%%%%%%%%%%%%%%%%%%%%%%%%%%%%%%%%%%%%%%%%%%%%%%%
% ---INTRODUCTION-------------------------------------------------
The promise of printable, flexible electronic devices and displays has fuelled the
development of the polymer field-effect transistor (pFET) over the past decade. However,
the long-term performance of state-of-the-art pFETs is limited by degradation mechanisms
that cause the threshold voltage to increase in
magnitude.\cite{Salleo03-142,Street03-145,Salleo04-190,Salleo05-141,advMat17_2411,Pernstich04-192}
The principal effect is thought to be charge carrier trapping in either the organic
semiconductor or at the semiconductor/insulator interface, which screens the applied gate
voltage. In many structures the effect of contact resistance on device degradation needs
to be considered, and can make the reliable extraction of the trapped-charge density
solely from $I-V$ characteristics a difficult task.\cite{Burgi03-194} It is therefore
desirable to use a noncontact technique, such as spectroscopy or potentiometry, to
investigate charge trapping in pFETs. In this letter, we report on a noncontact study of
the degradation mechanisms in polymer FETs, performed using a charge modulation technique
based on terahertz time-domain spectroscopy\cite{josab7_2006,cpl03} (TDS). Terahertz
radiation interacts strongly with charge carriers in a material, with a fractional
transmission change \DTonT (on injection or photoexcitation of charges) proportional to
the complex conductivity of the thin film.\cite{Lui01-189} We demonstrate that terahertz
TDS permits us to monitor the density of trapped holes in the accumulation layer by
coupling the low-mobility holes to higher-mobility electrons in the silicon gate.
Correlation of these findings with scanning probe potentiometry measurements allows us to
assess separately the contributions to transistor degradation arising from changes in the
contact resistance, field-effect mobility and trapped-carrier density.

%Terahertz TDS has been applied previously to photoexcited inorganic\cite{nat414_286} and
%organic\cite{prl92_196601} semiconductors, and superconductors.\cite{jap70_2238}

\begin{figure}[tb]
    \centering
    \includegraphics[width=8cm]{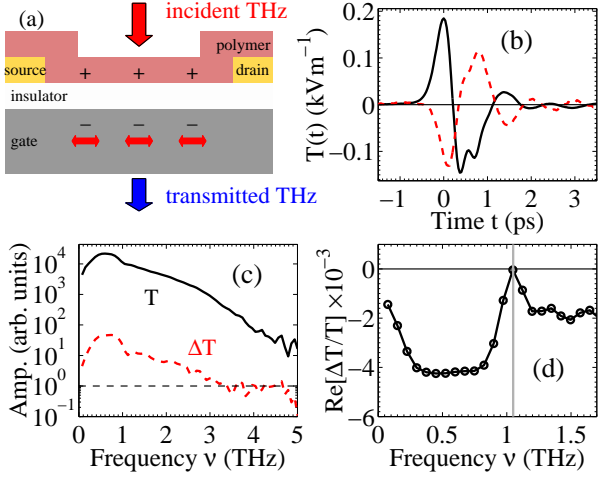}
    \caption{\label{FIG:geometry_THz} (Color online) (a) Schematic
    diagram of the transistor sample geometry for terahertz TDS. Application
    of a negative gate voltage gives rise to holes in the accumulation
    layer in the polymer and highly-mobile electrons in the silicon
    gate.  (b) Measured terahertz electric field $T(t)$ after transmission
    through transistor (solid line) and change in transmitted terahertz
    electric field ${\Delta}T(t)$ [dashed line, multiplied by
    $\times100$] upon application of a gate voltage
    $\Vgm=0\leftrightarrow-30$\,V. Both are given as a function of
    electro-optic sampling delay time $t$.  (c) Amplitude spectra of
    the transmitted THz radiation $T(\nu)$, and change in transmitted
    amplitude $\DTm(\nu)=T(\Vgm=-30$\,V$)-T(\Vgm=0$\,V$)$ obtained
    from the time-domain data in b) through Fourier transformation.
    (d) Change in transmission $\Delta T/T$ as a function of
    frequency. The artefact close to 1\,THz is a result of diffraction
    from the interdigitated array formed by the electrodes, as
    explained in the text.  }
\end{figure}

% %%%%%%%%% PFET details
A schematic diagram of the bottom-gate, bottom-contact polymer transistors fabricated for
this study is shown in Fig.\ \ref{FIG:geometry_THz}. The semiconducting polymer
poly[(9,9-dioctylfluorene-2,7-diyl)-co-(bithiophene)] (F8T2) was deposited through spin
casting from solution (in a layer of 100\,nm thick) onto an interdigitated gold array
(channel length of 40\,$\mu$m, finger width of 50\,$\mu$m, total channel width of
$45$\,mm). The gate electrode comprised a lightly $n$-doped silicon wafer
($2.5\times10^{15}$\,cm$^{-3}$) with a total thickness of 0.62\,mm, and a 200\,nm thick
SiO$_2$ gate dielectric. The transistors exhibited $p$-type conduction upon application
of a negative gate voltage and a source-drain bias.

% %%%%%%%%% THz-TDS setup
We used a terahertz time-domain spectrometer similar to the one described in
Ref.~\onlinecite{cpl03} to measure the terahertz radiation transmitted through the
transistor.
%A semi-insulating GaAs photoconductive switch biased with a 20\,kHz square
%wave at $\pm150$\,V generated a single-cycle electric-field transient after
%photoexcitation by pulses from a Ti:sapphire oscillator laser (10\,fs, 75\,MHz repetition
%rate, 400\,mW beam power). Electro-optic sampling was used to detect the transmitted
%transient.
To create charge-modulation effects, an a.c.\ square wave bias voltage \Vg was applied to
the gate, typically $\Vgm=0\leftrightarrow-30$\,V at $40$\,Hz.\cite{periodNote}  A
lock-in amplifier was used to measure the change \DT (resulting from the \Vg modulation)
in the terahertz electric field \T transmitted through the transistors [Fig.\
\ref{FIG:geometry_THz}(b)]. The terahertz beam and transistor were kept in a vacuum of
1\,mbar to minimize terahertz absorption from atmospheric water vapor.  In order to
obtain good transmission ($25$\,\%) through the device it was necessary to orient the
transistor with the fingers of the interdigitated array at $90$\degr to the plane of
polarization of the incident terahertz electric field. However, in this geometry the
interdigitated array diffracts the incident terahertz radiation at wavelengths close to
the repeat period of the array ($\lambda=90$\,$\mu$m in silicon, corresponding to
0.98\,THz in vacuum). This results in a first diffraction minimum in \DTonT near 1\,THz
[Figs.\ \ref{FIG:geometry_THz}c and d] and further reductions at higher frequencies. We
have therefore limited our data analysis to the unaffected free spectral range up to
$\sim$1\,THz.

% %%%%%%%%% SPKM
%As an alternative experimental approach, the local surface potential on operating
%transistors was measured via noncontact potentiometry, using a scanning force microscope
%in Kelvin probe mode. Here, the voltage applied to the conducting tip is regulated by a
%feedback loop to minimize the electrostatic force between tip and sample, with the
%resulting tip potential following the electrostatic potential in the accumulation layer.
%A detailed description of the experimental setup and the methods employed for data
%analysis and processing can be found in Ref.\cite{synMet146_297}.

% %%%%%%%%% RESULTS
Figs.\ \ref{FIG:geometry_THz}(b) and (c) display the measured change in terahertz
electric field $\Delta T$ under the application of a bias $\Vgm=0\leftrightarrow-30$\,V,
which is approximately 250 times smaller than the size of the electric field $T$
transmitted through the transistor. No transmission changes were observed after the F8T2
layer was chemically removed, or for devices fabricated without the polymer layer. Figure
\ref{FIG:geometry_THz}d shows that on the application of a gate bias the transistor
transmits less terahertz radiation (negative \DTonTn), indicating the creation of a
partially reflective layer through changes in the charge carrier density. A negative \Vg
induces both a hole accumulation layer on the polymer/insulator boundary, and an electron
accumulation layer of equal surface carrier density on the insulator/gate boundary [Fig.\
\ref{FIG:geometry_THz}(a)]. The observed transmission change of terahertz radiation with
applied gate bias arises primarily from the electrons in the silicon, as the mobility of
holes in the accumulation layer ($\mu=7\times10^{-3}$\,\cmVsn, estimated from the current
in the saturation regime) is more than five orders of magnitude lower than the electron
mobility in the silicon gate ($\sim 1400$\,\cmVsn). This interpretation is confirmed by
the lack of a transmission change found for an all-polymer transistor (on a quartz
substrate) within the experimental noise floor limit of $\DTonTm < 1\times 10^{-5}$. For
silicon-gate polymer transistors the electron layer in the gate therefore acts as an
indirect, but sensitive probe of the hole density in the polymer, by coupling it to
higher mobility electrons in the silicon.

\begin{figure}[bt]
    \centering
    \includegraphics[width=8cm]{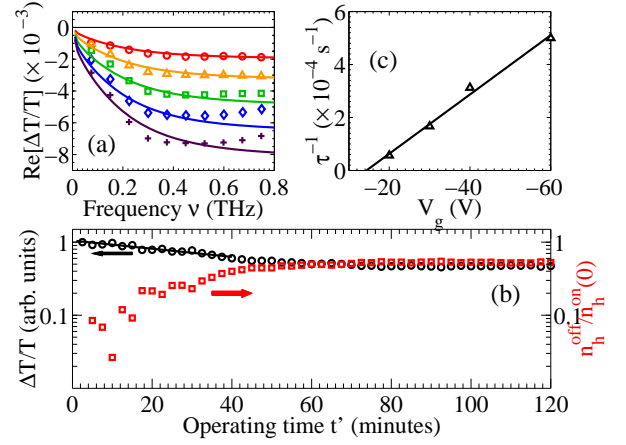}
    \caption{\label{FIG:bias-stress} (Color online) (a) Measured
    differential transmission \DTonT for a range of gate
    voltage modulations $0\leftrightarrow\Vgm$ with $\Vgm=$-10\,V
    (circles), -20\,V (triangles), -30\,V (squares), -40\,V
    (diamonds), and -50\,V (crosses). The solid lines are fits to the
    data based on the Drude-Lorentz thin-film model described in the
    text, with $\DTonTm \propto V_{\rm g} \propto n_{\rm h}$.  (b) When
    $\Vgm=0\leftrightarrow-30$\,V is applied for times $t'$ (half the
    measurement time, owing to the 50\% \Vg duty cycle) the
    differential transmission (circles) decays initially exponentially
    (straight line). The time constant $\tau=7.1\pm2.4 \times
    10^{3}$\,s of this decay was determined by averaging fits for
    three nominally identical transistors, over the first 40\,minutes.
    The trapped charge density remaining in the channel during the off
    period, $n^{\rm{off}}_{\rm{h}}(t')/n_{\rm h}^{\rm{on}}(0)$
    [squares, extracted from the modelled fits to \DTonTn, with charge
    density $n_{\rm h}^{\rm{on}}(0)=3.2\times 10^{12}$\,cm$^{-3}$ when
    on] saturates at large $t'$.  (c) Hole trapping rate $1/\tau$
    obtained from exponential fits to \DTonT during application of
    $0\leftrightarrow-\Vgm$ for 30\,min.  }
\end{figure}

\DTonT increases linearly with the applied gate voltage [Fig.\ \ref{FIG:bias-stress}(a)],
in accordance with an increase of charge density in the channel. We have modelled these
data using standard thin-film transmission coefficients. The accumulation layer in the
silicon gate was assumed to have a constant electron concentration $N_{\rm e}$ over a
thickness $\delta_{\rm e}$ at each gate voltage. The parameters used were in good accord
with those obtained from an analytical solution of Poisson's equation at the SiO$_2$/Si
boundary. The Drude-Lorentz model was used to calculate the dielectric function of the
electron layer, with scattering rate $\Gamma=1.5\times10^{12}$\,s$^{-1}$.

Excellent agreement with the experimentally measured \DTonT is obtained for an electron
accumulation layer density of $N_e^{\rm{off}}=2.5\times 10^{15}$\,cm$^{-3}$ in the `off'
($\Vgm=0$\,V) state and $N_{\rm e}^{\rm{on}}=4.0\times 10^{18}$\,cm$^{-3}$ in the `on'
($\Vgm=-30$\,V) state when $\delta_{\rm e}=8$\,nm, as shown in Fig.\
\ref{FIG:bias-stress}(a). The plasma frequency in the ``off'' (``on'') state is 0.2\,THz
(8.8\,THz). Assuming that the sheet charge density in the polymer ($n_h$) is the same as
that in the gate ($n_e$), the hole accumulation layer charge density for a pristine
transistor in the ``on'' state, $N_{\rm h}^{\rm{on}}$, can be calculated from $N_{\rm
h}^{\rm{on}}=N_{\rm e}^{\rm{on}}\delta_{\rm e}/\delta_{\rm h}$, where $\delta_{\rm h}$
and $\delta_{\rm e}$ are the thickness of the hole and the electron accumulation layer,
respectively. Taking $\delta_{\rm h}=1$\,nm as a reasonable approximation we obtain
$N_{\rm h}^{\rm{on}}=3.2\times 10^{19}$\,cm$^{-3}$ at $\Vgm=0\leftrightarrow-30$\,V, in
good agreement with typical values found in the literature.\cite{orgEl4_33} Using
identical parameter values to those determined above for $\Vgm=-30$\,V, but scaling
$n_{\rm e}$ and $\delta_{\rm e}$ linearly with gate voltage, results in model curves
closely matching the measured \DTonT over the entire range of applied \Vg [Fig.\
\ref{FIG:bias-stress}(a)].

The sensitivity of our technique to the hole density in the transistor channel makes it
an ideal tool to investigate the mechanisms governing degradation of these devices under
prolonged application of a gate bias voltage. Measurements of \DTonT as a function of
biasing time [given in Fig.\ \ref{FIG:bias-stress}(b)] show an exponential decrease for
approximately the first hour, after which the values gradually saturate. We attribute
these changes to an increase in density of trapped holes at the polymer/insulator
interface with time, resulting in a larger hole density $n_{\rm h}^{\rm{off}}$ in the
``off'' state (and therefore also an increased $n_{\rm e}^{\rm{off}}$, since trapped
holes remain in the channel and contribute to the signal). Figure
\ref{FIG:bias-stress}(b) displays $n_{\rm{h}}^{\rm{off}}$ as a function of operating
time, as extracted from the data using the model described above under the assumption
that all other parameters are unaffected by degradation. The hole density for the ``off''
state increases considerably within the first hour, but then saturates at a value of
approximately half that of the initial value in the ``on'' state $n_{\rm
h}^{\rm{on}}(0)$.

The exponential nature of the initial decay of the \DTonT signal indicates that the hole
trapping rate in the polymer is a linear function of the carrier density (i.e.\
$dn/dt=-n/\tau$) and therefore incompatible with the bipolaronic trapping mechanism
($dn/dt\propto n^2$) that has recently been proposed as a contributor to device
degradation on timescales below 1\,s.\cite{Salleo04-190,Street03-145} Fig.\
\ref{FIG:bias-stress}c demonstrates that the initial trapping rate $1/\tau$, extracted
from exponential fits to the initial decay, is proportional to the applied gate voltage.
This linear rise in $1/\tau$ suggests that the trapping cross-section or the trap density
(or both) increase with gate bias, as suggested recently by Salleo and
Street.\cite{Salleo04-190} We find that the decrease in \DTonT is temporarily reversible
under illumination, but only for photon energies above the polymer bandgap, confirming
that the degradation mechanism is largely associated with changes in the
polymer.\cite{synMet146_297} Similar device recovery is also found after leaving the
device with $\Vgm=0$\,V in the dark, as observed previously.\cite{Salleo03-142} The
modulation period used was too short to obtain significant carrier detrapping during the
`off' state, and consequently we attribute the trapping dynamics to a combination of
shallow and deep traps.

%%%%%%%%% SKPM
\begin{figure}[tb]
    \centering
    \includegraphics[width=8cm]{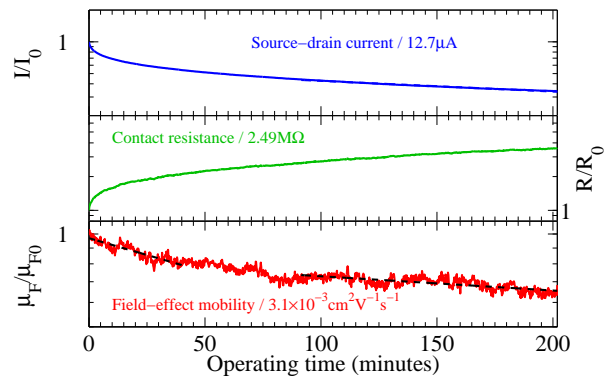}
    \caption{\label{FIG:SKPM} (Color online) Source current (top),
    source contact resistance (middle) and field-effect mobility
    $\mu_{\rm{F}}$ (bottom) of a 2\,$\mu$m channel length F8T2/300\,nm
    SiO$_2$/Si pFET as a function of operating time, normalized to
    their initial values and shown on a semi-logarithmic plot.  The
    curves were extracted from scanning Kelvin-probe microscopy
    measurements across the transistor channel. A constant gate
    voltage bias of $\Vgm=-40$\,V was applied, to produce an initial
    sheet charge density $n_h=2.9\times 10^{12}$\,cm$^{-2}$ comparable
    to that in the terahertz experiments.  The dotted lines are exponential
    fits to $\mu_{\rm{F}}$ at early and late operating times, with
    time constants $\tau=1.2 \times 10^{4}$\,s and $\tau=5.6 \times
    10^{4}$\,s respectively.}
\end{figure}
Finally, we compare the insights gained about polymer transistor degradation from
terahertz TDS techniques with those that may be obtained from more established techniques
based on noncontact potentiometry.\cite{Burgi02-195,Palermo06-146} For this purpose, we
have conducted scanning Kelvin-probe microscopy (SKPM) measurements, which can track the
electrostatic potential in the accumulation layer with a high spatial resolution
($<100$\,nm). The F8T2 transistors investigated were similar to those examined using
terahertz TDS apart from a reduced channel length (2\,$\mu$m), limited by the range of
the SKPM tip. Figure \ref{FIG:SKPM} displays the resulting source-drain current $I$,
contact resistance $R$, and channel field-effect mobility $\mu_{\rm F}=\mu_{\rm h} n_{\rm
h}$ as a function of operating time of the F8T2 transistor. It can be seen that the early
nonexponential decay of $I$ is caused by a rapid initial increase of the contact
resistance with operating time. The field-effect mobility, on the other hand, shows an
initial exponential decay ($\tau=1.2 \times 10^{4}$\,s) comparable to that obtained from
THz-TDS ($\tau=7.1 \times 10^{3}$\,s), before tending to saturate at longer operating
times. These results demonstrate the difficulty in extracting meaningful information
about the dynamics of carrier trapping in pFETs from $I-V$ characteristics, which are
significantly influenced by changes in contact resistance.\cite{Burgi03-194} The observed
decay of the field-effect mobility may be caused either by a decrease in hole mobility or
hole density in the channel. The combined THz-TDS and SKPM measurements therefore suggest
that the changes in field-effect mobility with transistor operation time are dominated by
a reduction in the density of mobile carriers, rather than a decrease in general mobility
of all charges in the channel.

In conclusion, we have investigated the mechanisms for degradation of polymer-based FETs
using a combination of terahertz spectroscopy and noncontact potentiometry.  The observed
terahertz transmission change under a gate bias were attributed to the layer of
high-mobility electrons that forms in the silicon gate as mirror charges to the
lower-mobility hole accumulation layer in the polymer. During the ``on'' state of the
transistor, the plasma frequency of the electron layer is shifted upwards in frequency,
permitting highly sensitive, noncontact probes of the accumulated charge density through
terahertz TDS. Our measurements demonstrate an initial monoexponential decrease of the
terahertz differential transmission signal with biasing time, in agreement with an
increase of trapped charge density in the polymer present also during the ``off'' state.
Complementary SKPM measurements show that the contact resistance strongly influences the
source-drain current at early device operation times ($<$40min). $I$-$V$ curves taken on
their own therefore do not provide direct access to the charge-trapping dynamics in
pFETs. By being sensitive only to electrons in the silicon gate, the terahertz TDS
measurements are not influenced by the hole mobility in the polymer. From the results of
both terahertz TDS and SKPM techniques we infer that an increase in trapped-charge
density, rather than a decrease in single-carrier mobility, is responsible for the
decline in field-effect mobility with operation time.

% %%%%%%%%% ACKNOWLEDGEMENTS
The authors would like to acknowledge support by the EPSRC for this work.

% %%%%%%%%% BIBLIOGRAPHY
%\bibliographystyle{apsrevJames}
%\bibliography{c:/thz/Reports/thzbib}

\begin{thebibliography}{21}
\expandafter\ifx\csname natexlab\endcsname\relax\def\natexlab#1{#1}\fi
\expandafter\ifx\csname bibnamefont\endcsname\relax
  \def\bibnamefont#1{#1}\fi
\expandafter\ifx\csname bibfnamefont\endcsname\relax
  \def\bibfnamefont#1{#1}\fi
\expandafter\ifx\csname citenamefont\endcsname\relax
  \def\citenamefont#1{#1}\fi
\expandafter\ifx\csname url\endcsname\relax
  \def\url#1{\texttt{#1}}\fi
\expandafter\ifx\csname urlprefix\endcsname\relax\def\urlprefix{URL }\fi
\providecommand{\bibinfo}[2]{#2} \providecommand{\eprint}[2][]{\url{#2}}

\bibitem[{\citenamefont{Sirringhaus}(2005)}]{advMat17_2411}
\bibinfo{author}{\bibfnamefont{H.}~\bibnamefont{Sirringhaus}},
  \bibinfo{journal}{Adv. Mat.} \textbf{\bibinfo{volume}{17}},
  \bibinfo{pages}{2411} (\bibinfo{year}{2005}).

\bibitem[{\citenamefont{Salleo and Street}(2003)}]{Salleo03-142}
\bibinfo{author}{\bibfnamefont{A.}~\bibnamefont{Salleo}} \bibnamefont{and}
  \bibinfo{author}{\bibfnamefont{R.~A.} \bibnamefont{Street}},
  \bibinfo{journal}{J. Appl. Phys.} \textbf{\bibinfo{volume}{94}},
  \bibinfo{pages}{471} (\bibinfo{year}{2003}).

\bibitem[{\citenamefont{Street et~al.}(2003)\citenamefont{Street, Salleo, and
  Chabinyc}}]{Street03-145}
\bibinfo{author}{\bibfnamefont{R.~A.} \bibnamefont{Street}},
  \bibinfo{author}{\bibfnamefont{A.}~\bibnamefont{Salleo}}, \bibnamefont{and}
  \bibinfo{author}{\bibfnamefont{M.~L.} \bibnamefont{Chabinyc}},
  \bibinfo{journal}{Phys. Rev. B} \textbf{\bibinfo{volume}{68}},
  \bibinfo{pages}{085316} (\bibinfo{year}{2003}).

\bibitem[{\citenamefont{Salleo and Street}(2004)}]{Salleo04-190}
\bibinfo{author}{\bibfnamefont{A.}~\bibnamefont{Salleo}} \bibnamefont{and}
  \bibinfo{author}{\bibfnamefont{R.~A.} \bibnamefont{Street}},
  \bibinfo{journal}{Phys. Rev. B} \textbf{\bibinfo{volume}{70}},
  \bibinfo{pages}{235324} (\bibinfo{year}{2004}).

\bibitem[{\citenamefont{Salleo et~al.}(2005)\citenamefont{Salleo, Endicott, and
  Street}}]{Salleo05-141}
\bibinfo{author}{\bibfnamefont{A.}~\bibnamefont{Salleo}},
  \bibinfo{author}{\bibfnamefont{F.}~\bibnamefont{Endicott}}, \bibnamefont{and}
  \bibinfo{author}{\bibfnamefont{R.~A.} \bibnamefont{Street}},
  \bibinfo{journal}{Appl. Phys. Lett.} \textbf{\bibinfo{volume}{86}},
  \bibinfo{pages}{263505} (\bibinfo{year}{2005}).

\bibitem[{\citenamefont{Pernstich et~al.}(2004)\citenamefont{Pernstich, Haas,
  Oberhoff, Goldmann, Gundlach, Batlogg, Rashid, and
  Schitter}}]{Pernstich04-192}
\bibinfo{author}{\bibfnamefont{K.~P.} \bibnamefont{Pernstich}},
  \bibinfo{author}{\bibfnamefont{S.}~\bibnamefont{Haas}},
  \bibinfo{author}{\bibfnamefont{D.}~\bibnamefont{Oberhoff}},
  \bibinfo{author}{\bibfnamefont{C.}~\bibnamefont{Goldmann}},
  \bibinfo{author}{\bibfnamefont{D.~J.} \bibnamefont{Gundlach}},
  \bibinfo{author}{\bibfnamefont{B.}~\bibnamefont{Batlogg}},
  \bibinfo{author}{\bibfnamefont{A.~N.} \bibnamefont{Rashid}},
  \bibnamefont{and} \bibinfo{author}{\bibfnamefont{G.}~\bibnamefont{Schitter}},
  \bibinfo{journal}{J. Appl. Phys.} \textbf{\bibinfo{volume}{96}},
  \bibinfo{pages}{6431} (\bibinfo{year}{2004}).

\bibitem[{\citenamefont{Burgi et~al.}(2003{\natexlab{a}})\citenamefont{Burgi,
  Richards, Friend, and Sirringhaus}}]{Burgi03-194}
\bibinfo{author}{\bibfnamefont{L.}~\bibnamefont{Burgi}},
  \bibinfo{author}{\bibfnamefont{T.~J.} \bibnamefont{Richards}},
  \bibinfo{author}{\bibfnamefont{R.~H.} \bibnamefont{Friend}},
  \bibnamefont{and}
  \bibinfo{author}{\bibfnamefont{H.}~\bibnamefont{Sirringhaus}},
  \bibinfo{journal}{J. Appl. Phys.} \textbf{\bibinfo{volume}{94}},
  \bibinfo{pages}{6129} (\bibinfo{year}{2003}{\natexlab{a}}).

\bibitem[{\citenamefont{Johnston et~al.}(2003)\citenamefont{Johnston, Herz,
  Khan, K\"{o}hler, Davies, and Linfield}}]{cpl03}
\bibinfo{author}{\bibfnamefont{M.~B.} \bibnamefont{Johnston}},
  \bibinfo{author}{\bibfnamefont{L.~M.} \bibnamefont{Herz}},
  \bibinfo{author}{\bibfnamefont{A.~L.~T.} \bibnamefont{Khan}},
  \bibinfo{author}{\bibfnamefont{A.}~\bibnamefont{K\"{o}hler}},
  \bibinfo{author}{\bibfnamefont{A.~G.} \bibnamefont{Davies}},
  \bibnamefont{and} \bibinfo{author}{\bibfnamefont{E.~H.}
  \bibnamefont{Linfield}}, \bibinfo{journal}{Chem. Phys. Lett.}
  \textbf{\bibinfo{volume}{377}}, \bibinfo{pages}{256} (\bibinfo{year}{2003}).

\bibitem[{\citenamefont{Grischkowsky et~al.}(1990)\citenamefont{Grischkowsky,
  Keiding, van Exter, and Fattinger}}]{josab7_2006}
\bibinfo{author}{\bibfnamefont{D.}~\bibnamefont{Grischkowsky}},
  \bibinfo{author}{\bibfnamefont{S.}~\bibnamefont{Keiding}},
  \bibinfo{author}{\bibfnamefont{M.}~\bibnamefont{van Exter}},
  \bibnamefont{and}
  \bibinfo{author}{\bibfnamefont{C.}~\bibnamefont{Fattinger}},
  \bibinfo{journal}{J. Opt. Soc. Am. B} \textbf{\bibinfo{volume}{7}},
  \bibinfo{pages}{2006} (\bibinfo{year}{1990}).

\bibitem[{\citenamefont{Lui and Hegmann}(2001)}]{Lui01-189}
\bibinfo{author}{\bibfnamefont{K.~P.~H.} \bibnamefont{Lui}} \bibnamefont{and}
  \bibinfo{author}{\bibfnamefont{F.~A.} \bibnamefont{Hegmann}},
  \bibinfo{journal}{Appl. Phys. Lett.} \textbf{\bibinfo{volume}{78}},
  \bibinfo{pages}{3478} (\bibinfo{year}{2001}).

\bibitem[{SKP()}]{periodNote}
\bibinfo{note}{The source and drain contacts were connected to 0\,V, and no source-drain
current flowed. The modulation period (0.025\,s) was chosen to be significantly longer
than both the estimated channel formation time ($\sim 9$\,$\mu$s, Ref.\ 12) and the $RC$
time constant created by the contact resistance ($\sim 1$\,ms).}

\bibitem[{\citenamefont{Burgi et~al.}(2003{\natexlab{b}})\citenamefont{Burgi,
  Friend, and Sirringhaus}}]{apl82_1482}
\bibinfo{author}{\bibfnamefont{L.}~\bibnamefont{Burgi}},
  \bibinfo{author}{\bibfnamefont{R.~H.} \bibnamefont{Friend}},
  \bibnamefont{and}
  \bibinfo{author}{\bibfnamefont{H.}~\bibnamefont{Sirringhaus}},
  \bibinfo{journal}{Appl. Phys. Lett.} \textbf{\bibinfo{volume}{82}},
  \bibinfo{pages}{1482} (\bibinfo{year}{2003}{\natexlab{b}}).

\bibitem[{\citenamefont{Burgi et~al.}(2004)\citenamefont{Burgi, Richards,
  Chiesa, Friend, and Sirringhaus}}]{synMet146_297}
\bibinfo{author}{\bibfnamefont{L.}~\bibnamefont{Burgi}},
  \bibinfo{author}{\bibfnamefont{T.}~\bibnamefont{Richards}},
  \bibinfo{author}{\bibfnamefont{M.}~\bibnamefont{Chiesa}},
  \bibinfo{author}{\bibfnamefont{R.~H.} \bibnamefont{Friend}},
  \bibnamefont{and}
  \bibinfo{author}{\bibfnamefont{H.}~\bibnamefont{Sirringhaus}},
  \bibinfo{journal}{Synth. Met.} \textbf{\bibinfo{volume}{146}},
  \bibinfo{pages}{297} (\bibinfo{year}{2004}).

\bibitem[{\citenamefont{Tanase et~al.}(2003)\citenamefont{Tanase, Meijer, Blom,
  and de~Leeuw}}]{orgEl4_33}
\bibinfo{author}{\bibfnamefont{C.}~\bibnamefont{Tanase}},
  \bibinfo{author}{\bibfnamefont{E.~J.} \bibnamefont{Meijer}},
  \bibinfo{author}{\bibfnamefont{P.~W.~M.} \bibnamefont{Blom}},
  \bibnamefont{and} \bibinfo{author}{\bibfnamefont{D.~M.}
  \bibnamefont{de~Leeuw}}, \bibinfo{journal}{Org. El.}
  \textbf{\bibinfo{volume}{4}}, \bibinfo{pages}{33} (\bibinfo{year}{2003}).

\bibitem[{\citenamefont{Burgi et~al.}(2002)\citenamefont{Burgi, Sirringhaus,
  and Friend}}]{Burgi02-195}
\bibinfo{author}{\bibfnamefont{L.}~\bibnamefont{Burgi}},
  \bibinfo{author}{\bibfnamefont{H.}~\bibnamefont{Sirringhaus}},
  \bibnamefont{and} \bibinfo{author}{\bibfnamefont{R.~H.}
  \bibnamefont{Friend}}, \bibinfo{journal}{Appl. Phys. Lett.}
  \textbf{\bibinfo{volume}{80}}, \bibinfo{pages}{2913} (\bibinfo{year}{2002}).

\bibitem[{\citenamefont{Palermo et~al.}(2006)\citenamefont{Palermo, Palma, and
  Samori}}]{Palermo06-146}
\bibinfo{author}{\bibfnamefont{V.}~\bibnamefont{Palermo}},
  \bibinfo{author}{\bibfnamefont{M.}~\bibnamefont{Palma}}, \bibnamefont{and}
  \bibinfo{author}{\bibfnamefont{P.}~\bibnamefont{Samori}},
  \bibinfo{journal}{Adv. Mater.} \textbf{\bibinfo{volume}{18}},
  \bibinfo{pages}{145} (\bibinfo{year}{2006}).

\end{thebibliography}

%\newpage
%\clearpage

\end{document}